\documentclass[12pt]{article}
\begin{document}
\title{Triangular mass matrices for quarks and leptons}
\author{ Stanis\l aw Tatur
\\ Nicolaus Copernicus Astronomical Center,\\ Polish Academy of
Sciences,\\ Bartycka 18, 00-716 Warsaw, Poland. \\ \\ \and Jan
Bartelski\\ Institute of Theoretical Physics,\\ Faculty of
Physics, Warsaw University,\\ Ho$\dot{z}$a 69, 00-681 Warsaw,
Poland. \\ }
\date{}
\maketitle
\begin{abstract}
\noindent We assume that all quark and lepton $3 \times 3$ mass
matrices which appear in the standard model lagrangian  (after
spontaneous symmetry breaking) with neutrinos being treated as
Dirac particles have the triangular form. Such matrices have not
only less non-zero elements (tree of them are equal to zero) but
also lead to very asymmetrical decomposition into one diagonal and
two unitary matrices for quarks and leptons. We also assume that
unitary matrices which transform flavor into definite mass states
for right handed components (weakly non-interacting) in the same
weak isospin doublet are equal. Using all available experimental
data on quark and lepton masses and mixing angles, treating in the
universal way quarks and leptons, we determine the triangular mass
matrices for up and down type quarks, neutrinos and charged
leptons and as a consequence mixing matrices for left-handed and
right-handed components. As the result of the fit we get
predictions for the neutrino masses including smallest neutrino
mass. The calculations without CP violation and  with inclusion of
this effect in quark sector are also presented.
\end{abstract}

\newpage
We have not only  an information from the experiment about quark
masses and Cabibbo-Kobayashi-Maskawa (CKM) \cite{CKM} mixing
matrix for quarks \cite{pdg} but now (after recent neutrino
oscillation experiments \cite{osc}) one also knows two mass square
differences for neutrinos and with less precision
Maki-Nakagawa-Sakata (MNS) \cite{MNS} mixing matrix for leptons
\cite{val,VMNS}. Both mixing matrices (CKM and MNS) are products
of matrices for left handed components for type $u$ and type $d$
quarks and for charged leptons and neutrinos. Knowing these
matrices from experiment we do not know what is the connection
between left and right handed components of the flavor states and
states with definite mass for up, down quarks, neutrinos and
charged leptons. It was pointed out that this creates a problem
when looking for patterns of masses and symmetries in neutrino and
quark sectors \cite{fer,king}. On the other hand if one knew from
the lagrangian, after spontaneous symmetry breaking, mass matrices
(they seem to be basic quantities) for  quarks and  leptons one
could find by decomposition in two unitary matrices and diagonal
matrix all the interesting  mixing matrices. Then one is able to
calculate CKM and MNS mixing matrices.

The aim of our paper is, treating neutrinos as Dirac particles, to
give the explicit example of possible mass matrices for $u$, $d$,
$\nu$ and $l^{-}$ giving mixing for left handed and right handed
components between states with definite flavor and definite mass.
We will assume that the mass matrices for quarks and leptons
gotten from standard model after spontaneous symmetry breaking are
in form of triangular matrices. With additional assumptions
discussed later we determine them using available experimental
information on quark and lepton masses and CKM and MNS mixing
matrices. We consider at the begining case without CP violation
and then include this effect in the quark sector.

Let us  introduce the notation for quarks. The $SU(3)*SU(2)*U(1)$
gauge invariant Yukawa interactions for quarks are given by
\begin{equation}
-L_{Y}=\bar{Q}_{i} (Y_{u})_{ij} u_{Rj} H^{\dagger}+
\bar{Q}_{i}(Y_{d})_{ij}d_{Rj}H + h.c. ,
\end{equation}

\noindent where $Q_{i}$ denote the $SU(2)$ doublets of left-handed
quarks and $u_{R}$, $d_{R}$ are the right-handed up and down-type
quarks respectively. The Yukawa couplings $Y_{u}$ and $Y_{d}$ are
$3\times3$ matrices (i,j are the generation indices) and $H$ is
the $SU(2)$ doublet Higgs field. After the electroweak symmetry
breaking, these Yukawa interactions lead to the following quark
mass terms
\begin{eqnarray}
-L_{m}&=&\bar{u}_{Li}(M_{u})_{ij}u_{Rj}+ \bar{d}_{Li}(M_{d})_{ij}d
_{Rj}+ h.c., \\ (M_{u})_{ij}&=&(Y_{u})_{ij}v, \nonumber \\
(M_{d})_{ij}&=&(Y_{d})_{ij}v, \nonumber
\end{eqnarray}

where $v$ is vacuum expectation value of the neutral component of
the Higgs field $H$.

These mass matrices $M_{u}$ and $M_{d}$ can be diagonalized using
two unitary matrices $U$ and $V$
\begin{eqnarray}
M_{u}&=&U_{u}M_{u}^{D}V_{u}^{\dagger}, \\
M_{d}&=&U_{d}M_{d}^{D}V_{d}^{\dagger}.
\end{eqnarray}

The diagonal matrix elements (of $M_{u}^{D}$ and $M_{d}^{D}$)
correspond to the experimentally observed mass eigenvalues. The
matrices $U$ and $V$ describe mixing between states with definite
flavor $u$, $d$ and with definite mass $u'$, $d'$ for left handed
and right handed components
\begin{eqnarray}
u_{L}&=&U_{u}u'_{L}, \nonumber \\  u_{R}&=&V_{u}u'_{R},  \\
d_{L}&=&U_{d}d'_{L}, \nonumber \\  d_{R}&=&V_{d}d'_{R}. \nonumber
\end{eqnarray}

The generation mixing in the charged weak current after expressing
in terms of fields with definite mass is  described by the
Cabibbo-Kobayashi-Maskawa (CKM) matrix which consists of two
unitary matrices
\begin{equation}
V_{CKM}=U_{u}^{\dagger}U_{d}.
\end{equation}

 In the similar way, assuming that
neutrinos are traditional Dirac particles, we have mixing for the
charged weak current in the lepton sector described by
Maki-Nakagawa-Sakata (MNS) matrix

\begin{equation}
V_{MNS}=U_{l}^{\dagger}U_{\nu},
\end{equation}

\noindent with the relations for neutrino mass $M_{\nu}$ and
charged leptons $M_{l}$

\begin{eqnarray}
M_{\nu}&=&U_{\nu}M_{\nu}^{D}V_{\nu}^{\dagger}, \\
M_{l}&=&U_{l}M_{l}^{D}V_{l}^{\dagger}.
\end{eqnarray}

\noindent As before diagonal matrix elements of $M_{\nu}^{D}$ and
$M_{l}^{D}$ correspond to the experimentally observed mass
eigenvalues.

The present experimental values of the moduli of the quark and
leptonic mixing matrices are for example given by \cite{pdg}

\begin{equation}
|V_{CKM}| =\left( \begin{array}{ccc} 0.9739 \div 0.9751&0.221 \div
0.227&0.0029 \div 0.0045 \\ 0.221 \div 0.227&0.9730 \div
0.9744&0.039 \div 0.044 \\ 0.0048 \div 0.014&0.037 \div
0.043&0.9990 \div 0.9992 \end{array} \right)
\end{equation}

and \cite{VMNS} (similar results given also by \cite{val})

\begin{equation}
|V_{MNS}| =\left( \begin{array}{ccc} 0.79 \div 0.88&0.47 \div
0.61&<0.20 \\ 0.19 \div 0.52&0.42 \div 0.73&0.58 \div 0.82
\\ 0.20 \div 0.53&0.44 \div 0.74&0.56 \div 0.81
\end{array} \right).
\end{equation}

In some sense matrices $M_{u}$,  $M_{d}$,  $M_{\nu}$ and  $M_{l}$
are basic quantities because they appear in lagrangian after
symmetry breaking. If we knew the numerical values of their matrix
elements we could immediately write the decomposition in terms of
unitary (ortogonal if there is no CP violation) matrices as in
eqs.(3), (4), (8) and (9) and then we could calculate $V_{CKM}$
and $V_{MNS}$.

Let us start with some intuitive argument for considering
triangular matrices. One can expect that because the ratios of
masses for $u$-type quarks are bigger then for $d$-type quarks the
mixing for $u$ type quarks is smaller then for $d$ type, in other
words, that the $U_{u}$ matrix is closer to identity matrix then
$U_{d}$. That means in this case that $U_{d} \approx V_{CKM}$. If
we also assume that $V_{d} \approx I$ (i.e. $V_{d}$ is not very
different from unity matrix) in fact we look in this way for
asymmetrical decomposition for $M_{d}$ and we will get in this
crude approximation from eq.(3) that $M_{d} \approx
V_{CKM}M_{d}^{D}$. Substituting experimental values we will get
that  $M_{d}$ is very close to triangular matrix with matrix
elements above main diagonal (with very small compared with the
other matrix elements terms below main diagonal). It seems that in
this case triangular matrices could be a good starting point to
describe quark mass matrices. On the other hand we would like to
treat quarks and leptons in the unified way. We assume that
triangular matrices describe mass matrices not only for quarks but
also for neutrinos and charged leptons. In this way we have 3 zero
elements in the mass matrix, so in some sense we follow the papers
in which one postulates several zeros in the mass matrix
\cite{par}. However, we do that in the universal way for quarks
and leptons. The advantage is that lower or upper triangular
matrices with hierarchical diagonal elements can give very
asymmetric decomposition namely that mixing in $U$ and $V$
matrices is very different. When we assume matrices in triangular
form which describe mass matrices of quarks and leptons we have
lower and upper triangular matrices at our disposal. The second
assumption is that  mass matrices for the members of weak isospin
doublets have different triangular form. That means that when for
example $d$ quark mass matrix is upper triangular matrix (it has
non-zero matrix elements above main diagonal) than $u$ quark mass
matrix is lower triangular matrix (non-zero matrix elements below
main diagonal and zero in the upper part). This assumption is
connected with the fact that to produce mass of $d$ and $u$ (lower
and upper weak isodoublet components) with the Higgs mechanism we
have coupling with Higgs doublet or its hermitian conjugate. The
third assumption is connected with the question how to take into
account, treating mass matrices as the basic objects, that the
right handed components are non active in weak interactions. The
masses of particles in the weak isomultiplets are very different.
We make an assumption that $V$ matrices responsible for the mixing
of right handed components in the same weak isodoublet are equal.
That of course gives a very strong limitation on the possible mass
matrices. With these three assumptions we are able  to calculate
mass matrices. Let us  start with quarks. Instead to solve eq.(3)
and (4) for quarks we will numerically minimalize the function

\begin{eqnarray}
\chi^2&=&\sum_{ij}
\frac{((U_{u}^{\dagger}U_{d})_{ij}-V_{CKMij}^{exp})^2}{(\Delta
V_{CKMij}^{exp})^2} + \sum_{ij}
\frac{(V_{uij}-V_{dij})^2}{(0.01)^{2}}+ \sum_{i}
\frac{(m_{di}^{D}-m_{di}^{exp})^{2}}{(\Delta m_{di}^{exp})^{2}}+
\nonumber \\ &+& \sum_{i}
\frac{(m_{ui}^{D}-m_{ui}^{exp})^{2}}{(\Delta m_{ui}^{exp})^{2}}.
\end{eqnarray}

There are 6 free parameters (3 diagonal and 3 non-diagonal) of
upper triangular mass matrix $M_{d}$ and 6 parameters of lower
triangular mass matrix $M_{u}$. The values
$m_{di}^{exp}=(m_{d},m_{s},m_{b})$,
$m_{ui}^{exp}=(m_{u},m_{c},m_{t})$ and corresponding errors
$\Delta m_{di}^{exp}$, $\Delta m_{ui}^{exp}$ are taken from the
Particle Data Group \cite{pdg}. When we take $V_{CKM}^{exp}$ from
eq.(10) we would not have in calculations unitarity of
$V_{CKM}^{exp}$ with enough precision and that would result in
minimal value of $\chi^2$, so we will take elements of
$V_{CKM}^{exp}$ and later $V_{MNS}^{exp}$ expressed in terms of
angles \cite{ohl} to have unitarity satisfied to high degree and
then use calculated in this way matrix elements. For the elements
of mixing matrices $V$ we take the artificial error 0.01.

The minimalization of $\chi^2$ gives the following results
(element of mass matrices in $MeV$)

\begin{equation}
M_{d} =\left( \begin{array}{ccc} 6.16&23.56&14.82
\\0&102.41&177.90 \\0&0&4246.25
\end{array} \right)
\end{equation}

\begin{equation}
U_{d} =\left( \begin{array}{ccc} 0.9744&0.2250&0.0035
\\-0.2249&0.9735&0.0419 \\0.0060&-0.0416&0.9991
\end{array} \right)
\end{equation}

\begin{equation}
M_{d}^{D}=\left( \begin{array}{ccc}6&0&0
\\0&105&0 \\0&0&4250
\end{array} \right)
\end{equation}

\begin{equation}
V_{d}^{\dagger}=\left( \begin{array}{ccc} 0.9999&-0.0132&8.5
\times 10^{-6}
\\0.0132&0.9999&-0.0010 \\5.1
\times 10^{-6}&0.0010&1
\end{array} \right)
\end{equation}

and

\begin{equation}
M_{u} =\left( \begin{array}{ccc} 2.75&0&0
\\16.49&1249.89&0 \\0.79&183.07&178000
\end{array} \right)
\end{equation}

\begin{equation}
U_{u} =\left( \begin{array}{ccc} 1&2.9 \times 10^{-5}&6.9 \times
10^{-11}
\\-2.9 \times 10^{-5}&1&7.2 \times 10^{-6} \\1.4
\times 10^{-10}&-7.2 \times 10^{-6}&1
\end{array} \right)
\end{equation}

\begin{equation}
M_{u}^{D}=\left( \begin{array}{ccc}2.75&0
\\0&1250&0 \\0&0&178000
\end{array} \right)
\end{equation}

\begin{equation}
V_{u}^{\dagger}=V_{d}^{\dagger}.
\end{equation}

We see that diagonal matrix elements  in mass matrices are very
close to the "experimental" quark masses for $u$ and $d$ type
quarks. Mixing in $V_{d}$ is much smaller then in $U_{d}$ and for
$u$ quark mixing in $U_{u}$ is smaller then in $V_{u}$. The
non-diagonal matrix elements in mass matrices $M_{u}$ and $M_{d}$
are comparable what is a little bit surprising. With our
assumptions we have got the solution for $u$ and $d$-type quarks
in which $U_{u}$ is with high accuracy equal to identity matrix.
The $V_{CKM}$ matrix calculated from eq.(6) is  identical to
$U_{d}$ obtained in eq.(14).

We will repeat the  calculations for neutrinos and charged
leptons. We do not know the neutrino masses so it is not possible
to solve eqs.(8) and (9). To get approximate solutions of eqs.(8)
and (9) we will proceed as before and minimalize suitably modified
$\chi^{2}$ function. In the first two terms in eq.(12) quark
mixing matrices are substituted by lepton mixing matrices,
$V_{CKM}$ by $V_{MNS}$ and so on. In the expression for $\chi^{2}$
given in eq.(12)  instead of the third term that corresponds to
diagonal $d$ quark masses we will take

\begin{equation}
\frac{((m_{2 \nu}^{D})^2-(m_{1 \nu}^{D})^2-(\Delta
m_{21}^{2})^{exp})^2}{(\Delta (\Delta m_{21}^{2})^{exp}))^2}+
\frac{((m_{3 \nu }^{D})^2-(m_{1 \nu}^{D})^2-(\Delta
m_{31}^{2})^{exp})^2}{(\Delta (\Delta m_{31}^{2})^{exp}))^2}.
\end{equation}

Instead of  the forth term in eq.(12) for $u$ quark we take the
corresponding term for charged leptons. The experimental values
$(\Delta m_{21}^{2})^{exp}$, $(\Delta m_{31}^{2})^{exp}$ and
corresponding errors are taken from the fit in \cite{VMNS}.
Neutrino masses are measured in eV while all other masses in MeV.
In the case of neutrinos equations corresponding to eqs.(13) to
(20) are

\begin{equation}
M_{\nu} =\left( \begin{array}{ccc} 0.00240&0.00469&-0.00078
\\0&0.01120&0.03308 \\0&0&0.03359
\end{array} \right)
\end{equation}

\begin{equation}
U_{\nu} =\left( \begin{array}{ccc} 0.8394&0.5436&0.0005
\\-0.3767&0.5810&0.7215 \\0.3919&-0.6058&0.6924
\end{array} \right)
\end{equation}

\begin{equation}
M_{\nu}^{D}=\left( \begin{array}{ccc}0.00203&0&0
\\0&0.00928&0 \\0&0&0.04781
\end{array} \right)
\end{equation}

\begin{equation}
V_{\nu}^{\dagger}=\left( \begin{array}{ccc} 0.9901&-0.1383&0.0237
\\0.1403&0.9759&-0.1674 \\2.7 \times 10^{-5}&0.169&0.9856
\end{array} \right)
\end{equation}

and

\begin{equation}
M_{l} =\left( \begin{array}{ccc} 0.516&0&0
\\14.828&106.134&0 \\-0.096&299.318&1751.51
\end{array} \right)
\end{equation}

\begin{equation}
U_{l} =\left( \begin{array}{ccc} 1&0.0007&8.7 \times 10^{-9}
\\-0.0007&0.9999&0.0101 \\6.9
\times 10^{-6}&-0.0101&0.9999
\end{array} \right)
\end{equation}

\begin{equation}
M_{l}^{D}=\left( \begin{array}{ccc}0.511&0&0
\\0&105.658&0 \\0&0&1776.99
\end{array} \right)
\end{equation}

\begin{equation}
V_{l}^{\dagger}=V_{\nu}^{\dagger}.
\end{equation}

We  get that (non-diagonal) elements in mixing matrix $U_{\nu}$
are bigger then in $V_{\nu}$ and in $U_{l}$ are smaller then in
$V_{l}$. $U_{l}$ is not very different from the unit matrix. For
the  $V_{MNS}$ mixing matrix calculated according to eq.(7) we
get:
\begin{equation}
V_{MNS} =\left( \begin{array}{ccc} 0.8396&0.5432&0
\\-0.3800&0.5875&0.7245 \\0.3881&-0.6000&0.6997
\end{array} \right).
\end{equation}
The differences between $V_{MNS}$ and $U_{\nu}$ matrix elements
are smaller then experimental errors for matrix elements of
$V_{MNS}$. Non-diagonal matrix elements in mass matrix for charged
leptons are comparable with those for $u$ and $d$ quarks. This is
not a case for neutrino mass matrix. Matrix elements in $V_{\nu}$
and in $V_{l}$ are not small and comparable with Cabibbo mixing in
$d$-type quarks. This result coming from the fit and being a
consequence of our third assumption is very promising when we
think about some ratios of charged lepton masses as being
comparable in magnitude to the ratios of down quark masses. In the
neutrino mass matrix non-diagonal matrix elements are comparable
with the diagonal ones what produces strong mixing in $U_{\nu}$
and relatively strong in $V_{\nu}$. Diagonal matrix elements in
neutrino  mass matrix eq.(22) are different from the diagonal
masses in the matrix $M_{\nu}^{D}$ eq.(24) because of strong
mixing. According to eq.(24) we get:

\begin{eqnarray}
m_{1 \nu}&=&0.002\; eV, \nonumber \\m_{2 \nu}&=&0.009\; eV,
\\ m_{3 \nu}&=&0.048 \;eV. \nonumber
\end {eqnarray}

These values of neutrino masses come out from the fit. We get some
hierarchy in neutrino mases but not comparable with what we have
for quarks and charged leptons. Specially interesting is the first
mass eigenvalue because it sets a scale. Its mass is about 1/4 of
the mass of the second neutrino.The values of $m_{2 \nu}$ and
$m_{3 \nu}$ are close to what one would expect from mass squared
differences with small first neutrino mass. The question is what
is the error for $m_{1 \nu}$ in the fit. When we fix $m_{1 \nu}$
and make fit for other quantities we get minimal $\chi^{2}$
smaller then one for $0.002 \leq m_{1 \nu} \leq 0.005$. For much
smaller values of $m_{1 \nu}$ we also get $\chi_{min}^{2} \leq 1$
so we do not have strong limitations for $m_{1 \nu}$. It seems
that except of neutrino when mixing in mass matrix is very strong
and the scale of masses is completely different the non-diagonal
matrix elements in mass matrices of $u$ and $d$ type quarks and
charged leptons are comparable (not very different in magnitude)
in spite of the differences in ratios of masses of quarks and
charged leptons. Could it be a hint in understanding the structure
of masses of quarks and leptons?

Till now all the mixing matrices were real (orthogonal). There is
no problem with repeating the calculation taking into account CP
violation. In this case we simply have additional imaginary parts
in $V_{CKM}$ matrix and unitary instead orthogonal mixing
matrices. When we repeat the calculation as before for $u$ and $d$
type quarks solving eqs.(3) and (4) by minimizing $\chi^{2}$
function corresponding to that given in eq.(12) and taking into
account that we have complex quantities we get:

\begin{equation}
M_{d} =\left( \begin{array}{ccc} 6.16&23.55-0.01i&7.67-12.69i
\\0&102.41&177.90-0.002i \\0&0&4246.25
\end{array} \right)
\end{equation}

\begin{equation}
U_{d} =\left( \begin{array}{ccc} 0.9744&0.2250&0.0018-0.0030i
\\-0.2249-0.0001i&0.9735-3 \times 10^{-5}i&0.0419 \\
0.0077-0.0029i&-0.0412-0.0007i&0.9991
\end{array} \right)
\end{equation}

\begin{equation}
M_{d}^{D}=\left( \begin{array}{ccc}6&0&0
\\0&105&0 \\0&0&4250
\end{array} \right)
\end{equation}

\begin{equation}
V_{d}^{\dagger}=\left( \begin{array}{ccc}
0.9999&-0.0132&(1.1+0.4i) \times 10^{-5}
\\0.0132&0.9999&-0.0011+1.7 \times 10^{-5}i \\(0.3+0.4i)
\times 10^{-5}&0.0011+1.9 \times 10^{-5}i&1
\end{array} \right)
\end{equation}

and

\begin{equation}
M_{u} =\left( \begin{array}{ccc} 2.75&0&0
\\16.49-0.0002i&1249.89&0 \\0.47+0.77i&181.41+2.96i&178000
\end{array} \right)
\end{equation}

\begin{equation}
U_{u} =\left( \begin{array}{ccc} 1&2.9 \times 10^{-5}&(4.1 -6.7i)
\times 10^{-11}
\\-2.9 \times 10^{-5}&1&(7.8 -0.1i) \times 10^{-6}
\\(1.6-0.6i)
\times 10^{-10}&(-7.2 -0.1i) \times 10^{-6}&1
\end{array} \right)
\end{equation}

\begin{equation}
M_{u}^{D}=\left( \begin{array}{ccc}2.75&0&0
\\0&1250&0 \\0&0&178000
\end{array} \right)
\end{equation}

\begin{equation}
V_{u}^{\dagger}=V_{d}^{\dagger}.
\end{equation}

Eqs.(32) to (39) should be compared with those given in
eqs.(13-20). We see that there is no big change except of
additional imaginary parts. There is no problem with repeating
this type of calculation for neutrinos and charged leptons taking
into account CP violation but we do not have enough information
about $V_{MNS}$ matrix at the moment.

We have shown that it is possible to determine mass matrices for
quarks and leptons under the assumption that they are triangular
and that the matrices for upper and lower weak isospin components
have zeros below or above main diagonal. The mixing matrices
corresponding to diagonalization of weakly non-interacting right
handed components are assumed in weak isodoublet to be equal. As
an input the values of matrix elements for CKM and MNS matrices
with known errors and masses of quarks and charged leptons and in
the case of neutrinos squared mass differences also with
corresponding errors are used. The masses of neutrinos are
determined in the fit and the possible range of smallest neutrino
mass is given. The $U$ matrices for up type quarks and charged
leptons are not very different from unit matrix. The $V$ matrix
for down type quarks is close to unit matrix and for neutrino
matrix elements of mixing matrix are not small and comparable with
Cabibbo mixing. In this solution $U$ matrices for $d$ type quarks
and neutrinos are very close to measured $V_{CKM}$ and $U_{MNS}$
matrices. It was actually our aim to find out whether it is
possible to get this type of solution. For up, down and charged
leptons the non-diagonal matrix elements in the mass matrices are
of the same order what is not expected taking into account masses
of these particles (up and down type quarks and charged leptons).
Neutrino masses have different scale (the hierarchy is rather
weak) and some non-diagonal matrix elements are comparable with
the diagonal. In the case of up and down type quarks we also
calculated using our method mass matrices and mixing matrices
taking into account CP violating phase.


\newpage

\begin{thebibliography}{99}
\bibitem{CKM} N. Cabibbo,  Phys. Rev. {\bf 10}, 531 (1963);
M.Kobayashi, T.Masakawa Progr. Theor. Phys. {\bf 49}, 652 (1973);
\bibitem{pdg} S.Eidelman {\em et al.}
(Particle Data Group), Phys. Lett. B {\bf 592}, 1 (2004);
\bibitem{osc} Super-Kamiokande Y.Fukuda et. al. Phys. Rev. Lett. {\bf 81}, 1562 (19998);
SNO Ahmad {\em et al.} Phys. Rev. Lett. {\bf 87}, 071301 (2001);
SNO Ahmad {\em et al.} Phys. Rev. Lett. {\bf 89}, 011301 (2002);
KAMLAND K.Eguchi {\em et al.}  Phys. Rev. Lett. {\bf 90}, 021802
(2003); K2K M.H.Ahn {\em et al.}  Phys. Rev. Lett. {\bf 90},
041801 (2003);
\bibitem{MNS} Z.Maki, M.Nakagawa, S.Sakata Progr. Theor. Phys. {\bf 28}, 870 (1962)
\bibitem{val} M.Maltoni, T.Schweth, M.A.T\'{o}rtola, J.W.F.Valle
New J. Phys. {\bf 6}, 122 (2004); J.W.F.Valle hep-ph/0603223;
\bibitem{VMNS} M.C.Gonzalez-Garcia Phys. Scripta T{\bf 121}, 72 (2005)
\bibitem{fer} F.Ferugio Nucl. Phys. Proc. Suppl. {\bf 143}, 184
(2005); Nucl. Phys. Proc. Suppl. {\bf 143}, 225 (2005);
\bibitem{king} S.F.King Rept. Prog. Phys. {\bf 67}, 107 (2004);
\bibitem{par} N.Uekusa, A.Watanabe, K.Yoshioka  Phys. Rev. D {\bf 71}, 094024 (2005);
C.Hagedorn, W.Rodejohann JHEP {\bf 0507}, 034 (2005); S.Kaneko,
H.Sawanaka, M.Tanimoto PoHEP2005, 187 (2006);
\bibitem{ohl} T.Ohlsson Phys. Lett. B {\bf 622}, 159 (2005);

\end{thebibliography}
\end{document}